\documentclass[conference]{IEEEtran}
\IEEEoverridecommandlockouts
\usepackage{cite}
\usepackage{url}
\PassOptionsToPackage{hyphens}{url}\usepackage{hyperref}
\usepackage{amsmath,amssymb,amsfonts}
\usepackage{algorithmic}
\usepackage{graphicx}
\usepackage{textcomp}
\usepackage{xcolor}
\usepackage{float}
\usepackage{dblfloatfix}
\usepackage{booktabs}
\usepackage{lipsum,multicol}
\usepackage{arydshln}
\def\BibTeX{{\rm B\kern-.05em{\sc i\kern-.025em b}\kern-.08em
    T\kern-.1667em\lower.7ex\hbox{E}\kern-.125emX}}
\begin{document}

\title{Lead Sheet Generation and Arrangement by Conditional Generative Adversarial Network\\
% or
%\title{Phrase-level Lead Sheet Arrangement by Recurrent Convolutional Generative Adversarial Network\\
% or...
%\title{Phrase-level Lead Sheet Generation and Its Chord-conditioned Arrangement using Recurrent Convolutional Generative Adversarial Network\\
}

% \IEEEauthorblockA{\textit{Affiliation} \\
% %\textit{name of organization}\\
% \and
% \IEEEauthorblockA{\textit{Affiliation} \\

\author{\IEEEauthorblockN{Hao-Min Liu}
\IEEEauthorblockA{
\textit{Research Center for IT Innovation} \\
\textit{Academia Sinica}\\
Taipei, Taiwan \\
paul115236@citi.sinica.edu.tw}
\and
\IEEEauthorblockN{Yi-Hsuan Yang}
\IEEEauthorblockA{
\textit{Research Center for IT Innovation} \\
\textit{Academia Sinica}\\
Taipei, Taiwan \\
yang@citi.sinica.edu.tw}
}

\maketitle

\begin{abstract}
% This paper propose the first model for lead sheet arrangement. There are three contributions shown in this paper. First of all, a phrase-level lead sheet generation, using two-track sequential RCNN-generative model, is proposed. Second, a track-conditional generative model for multi-track polyphonic arrangement is proposed. Third, a new idea on combining lead sheets and MIDIs is proposed with its implementation through three different features (chroma piano-roll, chroma beats, chord piano-roll). Finally, we report objective and subjective evaluation on lead sheet generation model as well as arrangement generation model, respectively. The audio samples of both models can be found at \texttt{\url{https://drive.google.com/open?id=1bYehFrp_B9B1OiJO_0B6orlAvaXIX6B8}}\\
Research on automatic music generation has seen great progress due to the development of deep neural networks. However, the generation of multi-instrument music of arbitrary genres still remains a challenge. Existing research either works on lead sheets or multi-track piano-rolls found in MIDIs, but both musical notations have their limits.
In this work, we propose a new task called \emph{lead sheet arrangement} to avoid such limits. A new recurrent convolutional generative model for the task is proposed, along with three new symbolic-domain harmonic features to facilitate learning from unpaired lead sheets and MIDIs. Our model can generate lead sheets and their arrangements of eight-bar long. Audio samples of the generated result can be found at \texttt{\url{https://drive.google.com/open?id=1c0FfODTpudmLvuKBbc23VBCgQizY6-Rk}}\\

\end{abstract}


\section{Introduction}
Automatic music generation by machine (or A.I.) has regained great academic and public attention in recent years, due largely to the development of deep neural networks \cite{briot17survey}. Although this is not always made explicitly, researchers often assume that machine can learn the rules people use to compose music, or even create new rules, given a sufficient amount of training data (i.e. existing music) and a proper neural network architecture. Thanks to cumulative efforts in the research community, deep neural networks have been shown successful in generating monophonic melodies \cite{melodyrnn,folkrnn} or polyphonic music of certain musical genres such as the Bach Chorales \cite{bachbot,deepbach} and piano solo \cite{performanceRNN,huang18icmlw}.

%Through great success in image or text generation using several deep neural networks as generative models, music generation also becomes a hot topic in recent years. 

\begin{figure}[t]
\centering
\includegraphics[width=.4\textwidth]{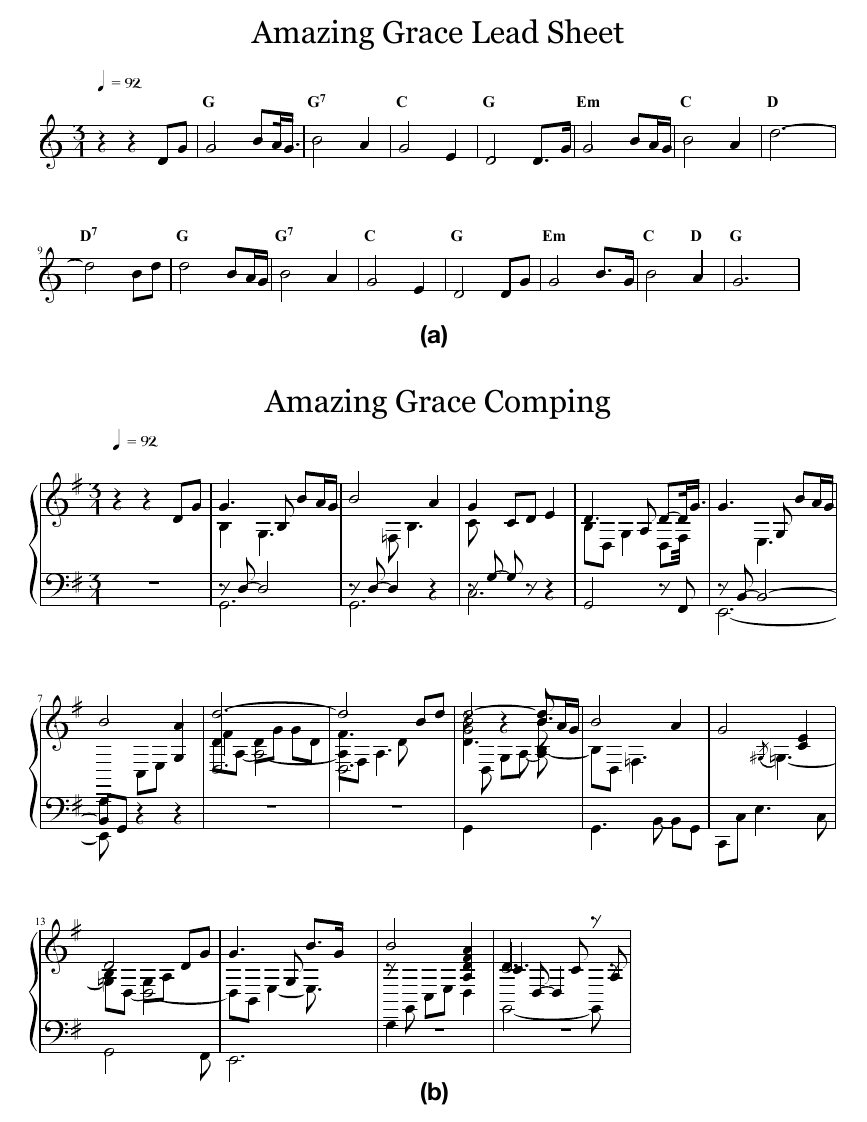}
\caption{(a) A lead sheet of \emph{Amazing Grace} and (b) its arrangement version.}
\label{fig:leadsheet_arr}
\end{figure}

One of the major remaining challenges is that of generating polyphonic music of arbitrary genres. 
Existing work can be divided into two groups, depending on the target form of musical notation. The first group of research aims at creating the \emph{lead sheet} \cite{midinet,lim17ismir,tsushima17ismir,roy17arxiv}, which is composed of a melody line and a sequence of accompanying chord labels (marked above the staff, see Fig. \ref{fig:leadsheet_arr}(a) for an example). %Compared to generating both the melody and chords from scratch, more research work has been done for melody harmonization, namely generating the chord sequence given a melody line.
A lead sheet can be viewed as a middle product of the music. In this form, the melody is clearly specified, but the accompaniment part is made up of only the chord labels. It does not describe the chord voicings, voice leading, bass line or other aspects of the accompaniment \cite{leadsheet_wiki}.
The way to play the chords (e.g. to play all the notes of a chord at the same time, or one at a time) is left to the interpretation of the performer.
%which only inform the harmonic information within certain duration. 
%but there is no performing information. On the one hand, 
%While a lead sheet gives the composers the freedom and flexibility to perform the music, it 
Accordingly, the rhythmic aspect of chords is missing.

The second group of research aims at creating music of the \emph{MIDI} format \cite{musegan,bmusegan,simon18arxiv}, which indicates all the voicing and accompaniment of different instruments. A MIDI can be represented by a number of piano-rolls (which can be seen as matrices) indicating the active notes per time step per instrument \cite{raffel16ismir,musegan}. For example, the MIDI for a song of a four-instrument rock band would have four piano-rolls, one for each instrument: lead guitar, rhythm guitar, bass, and drums. A MIDI contains detailed information regarding what to be played by each instrument. 
But, the major problem is that a MIDI typically does not specify which instrument plays the melody and which plays the chord.
%but lacks the information of melody line, which is the theme of the music. Also it does not provide the chord labels, which is a middle level information on the aspect of harmonicity.

\begin{figure*}[t]
\centering
\includegraphics[width=.9\textwidth]{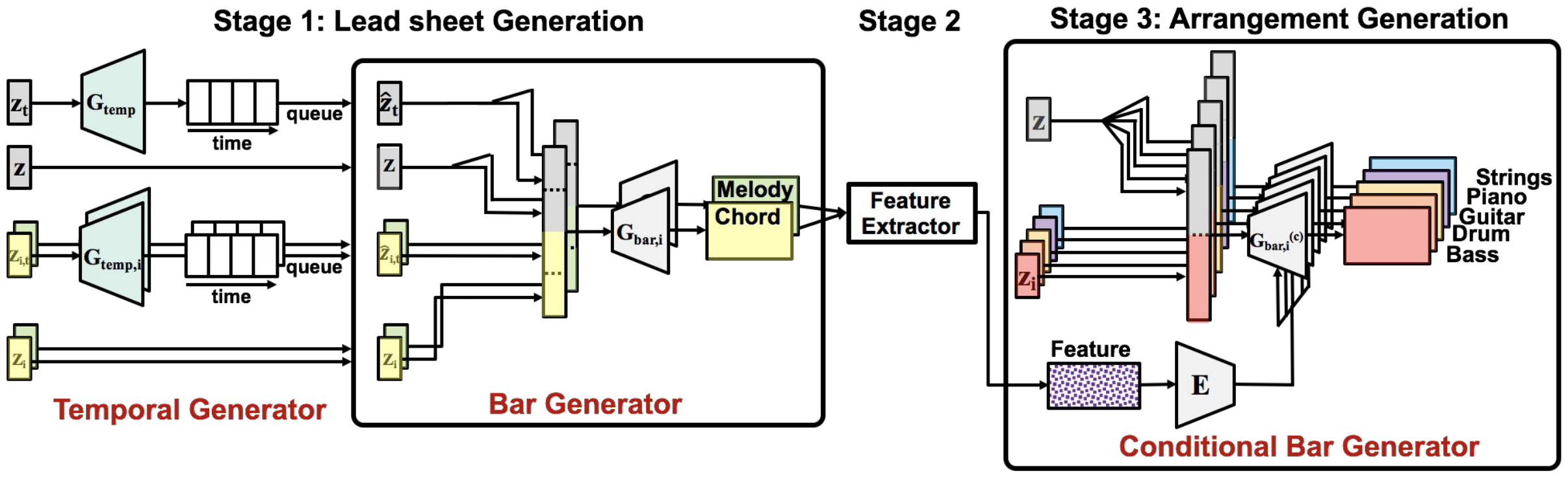}
\caption{Architecture of the proposed recurrent convolutional generative adversarial network model for lead sheet generation and arrangement.}
\label{fig:generator_arch}
\end{figure*}

An interesting and yet rarely-studied topic is the generation of something in the middle of the aforementioned two forms. We call it \emph{lead sheet arrangement}. 
%generate both lead sheet and its different kinds of arrangement. 
Arranging is the art of giving an existing melody musical variety \cite{arrangement}. It can be understood as the process to accompany a melody line or solo with other instruments based on the chord labels on the lead sheet \cite{leadsheet_wiki}. For example, Fig. \ref{fig:leadsheet_arr}(b) shows an arrangement of the lead sheet of the popular gospel song, \emph{Amazing Grace}.
We see that it shows how the chords are to be played.
%, which includes the chord voicing and comping.
%Figure \ref{fig:leadsheet_arr}(a) presents lead sheet version of this song, which includes melody line and the corresponding chord labels. Figure \ref{fig:leadsheet_arr}(b) presents the same song in midi version, which shows the arrangement, including voicing and comping of it. 
%Lead sheet arrangement is always critical for composers, or more specifically, for music arranger. 
%Melody line has a decisive influence in composing a song for certain; however, the arrangement also has a serious impact on the song performance. 
Arrangement can be seen across different genres. In classical music, arrangement is used in basic type such as sonata, string quartet and also more complex type as symphony. In Jazz, arrangement is often used to accompany the solo instrument. Pop music nowadays also have lots of arrangement setting to increase the diversity and richness of the music. 

Computationally, we define lead sheet arrangement as the process that takes as input a lead sheet and generates as output piano-rolls of a number of instruments to accompany the melody of the given lead sheet. In other words, the lead sheet is treated as a ``condition'' in generating the piano-rolls.
For example, we aim to generate the piano-rolls of the following instruments in our implementation: strings, piano, guitar, drum and bass. Compared to MIDIs, the result of lead sheet arrangement is musically more informative, for the melody line is made explicit.

To our knowledge, there is no prior work on lead sheet arrangement, possibly because of the lack of a dataset that contains both the lead sheet and MIDI versions of the same songs. Our work presents three technical contributions to circumvent this issue:
\begin{itemize}
\item First, we develop a new conditional generative model (based on generative adversarial network, or GAN \cite{gan}) 
%cGAN
to learn from unpaired lead sheet and MIDI datasets for lead sheet arrangement.
As Fig. \ref{fig:generator_arch} shows, the proposed model  contains three stages: \emph{lead sheet generation}, \emph{feature extraction}, and \emph{arrangement generation} (see Section \ref{sec:model} for details). The middle one  plays a pivotal role by extracting symbolic-domain harmonic features from the given lead sheet to condition the generation of the arrangement. For the conditional generation to be possible, we use features that can be computed from both the lead sheets and the MIDIs. 
\item Second, we propose and empirically compare the performance of three such harmonic features to connect lead sheets and MIDIs for lead sheet arrangement. They are \emph{chroma piano-roll} features, \emph{chroma beats} features, and \emph{chord piano-roll} features (see Section \ref{sec:model:flow}).
\item We employ the convolutional GAN model proposed by Dong \emph{et al.} for multi-track piano-roll generation \cite{musegan} for both lead sheet generation and arrangement generation, since the former can be viewed as two-track piano-roll generation and the later can be viewed as conditional five-track piano-roll generation. As a minor contribution, we replace the convolutional layers in the temporal generators (marked as $G_\text{temp}$ and $G_\text{temp,i}$ in Fig. \ref{fig:generator_arch}) by recurrent layers, to better capture the repetitive patterns seen in lead sheets. As a result, we can generate realistic lead sheets (and their arrangement) of eight bars long.
%architecture, composed by RNN temporal and CNN local generative model ,is proposed for long structured (eight bars) lead sheet music generation. Comparing to MuseGAN, using only CNN structure in all generative models, this architecture could capture the repetitive pattern shown in theorytab dataset.
%\item A track conditioned long structured arrangement generative model is proposed. To the best of our knowledge, it is the first long structured conditional model that can generate multitrack with either monophonic or polyphonic music in each track.
\end{itemize}

In our experiment, we evaluate the effectiveness of the lead sheet generation component and the overall lead sheet arrangement model through objective metrics and subjective evaluation, respectively. Upon publication, we will set up a github repo to demonstrate the generated music and to share the code for reproducibility.
%Unlike lead sheet generation or MIDI (multi-track piano-roll) generation, there is a lack of data for 
%Although this is really helpful and worth try but because of the lacking on dataset containing both lead sheet and midi information together, there is no paper implement lead sheet arrangement.

% Previous works could also be divided into these two categories according to the type of dataset they use. Leadsheet models take leadsheet, usually in xml form, as input while polyphonic and multi-track models take midi files as input. These two forms are quite different and each of them has its own advantages and limitations. Lead sheet is constituted by a melody line and its harmonic part, which is specified with chord labels. Therefore, it explicitly denotes where the melody is. However, it lacks the information of how the chord label should be played. Midi representation, on the other hand, shows all the voicing and accompaniment among different instruments but lacks of the information of melody line and chord label. Due to the shortage on the dataset, which contains melody, chord label with its arrangement, we, proposed a model that links lead sheet and midi dataset, and realize lead sheet arrangement in this paper.

\section{Related Work}
Many deep learning models have been proposed for lead sheet generation, possibly because lead sheets are commonly used in pop music. 
Lead sheet generation can be done in three ways: given chords, generate melody (e.g., \cite{midinet}); given melody, generate chords (a.k.a. melody harmonization, e.g., \cite{lim17ismir,tsushima17ismir}); or generating both melody and chords from scratch (e.g., \cite{roy17arxiv}). 
Some recent models generate not only the melody and chords but also the drums \cite{songfrompi}. But, as the target form of musical notation is still the lead sheet, only a sequence of chord labels (usually one chord label per bar or per half-bar) is generated, not the chord voicing and comping \cite{comping}.

%MelodyRNN model~~\cite{melodyrnn}, proposed by Google Magenta, aims at generating melody based on a chord sequence. In order to yield longer structure, they also implement its variants, LookbackRNN and AttentionRNN. MidiNet\cite{midinet} is another chord conditioned melody generation model. However, they use CNN with a generative adversarial network (GAN) structure to produce lead sheet. Semi-rnn cnn-based vae-GAN model\cite{crnngan} is also proposed for melody generation. It is .... 

%Among all previous works, we found that there are two parts that could be improved. Firstly, instead of using a given chord sequence as the generation condition, we could generate melody and chord progression either from scratch or conditional mode. Second, rather than conditioning the whole bar on just one chord, we generate chord pattern in with unconstrained rhythmic pattern using piano-roll representation.

%Besides lead sheet generation, there are also several works starting to focus on polyphonic and multi-track music generation. Song From PI\cite{songfrompi} proposed a model which generates pop song music with multiple tracks. First comes the melody, then the chord and drum are both generated based on the melody. Moreover, chord label and drum pattern are both generated in half-bar scale. 

To generate music that uses more instruments/tracks, Dong \emph{et al.} proposed MuseGAN, a convolutional GAN model that learns from MIDIs for multi-track piano-roll generation \cite{musegan}. MuseGAN creates music of five tracks (strings, piano, guitar, drum and bass), whereas a later variant of it that uses binary neurons considers eight tracks (replacing strings by ensemble, reed, synth lead and synth pad) \cite{bmusegan}. 
Recently, Simon \emph{et al.} \cite{simon18arxiv} used a variational auto-encoder (VAE) model to obtain a bar embedding space with instrumentation disentanglement and a chord conditioned structure. Their model demonstrates the ability of generating polyphonic measures conditioned on predefined chord progression. The instrumentation in this model can be viewed as part of arrangement. Another work that is concerned with the instrumentation of music is the MIDI-VAE model proposed by Brunner \emph{et al.} \cite{midivae}. Learning from MIDIs, the model creates a bar embedding space with instrument, pitch, velocity disentanglement and style label condition. Although all these models can arrange the instruments in multi-track music, they cannot specify which track plays the melody. % and which plays the chords.

There are some other models for multi-track music generation. % that do not use piano-rolls. 
DeepBach \cite{deepbach} aims at generating four-track chorales in the style of Bach. 
%model produces polyphonic music and hymn-like pieces by using Chain Monte Carlo (MCMC) algorithm. 
Jambot \cite{jambot} generates a chord sequence first and then uses that as a condition to generate polyphonic music of a single track (e.g., piano solo).  
%also proposes an approach to achieve polyphonic music generation. They predict a chord progression from a chord embedding by a LSTM model. Then, they generate polyphonic music based on the predicted chord progression by a second LSTM model. Some music theory concepts are matched with their learned chord embeddings. 
MusicVAE \cite{musicvae}, an improved version of Google Magenta's MelodyRNN model \cite{melodyrnn}, uses a hierarchical autoencoder to create melody, bass and drums  (i.e. three tracks) of 32 bars long. However, all these models are based on recurrent neural networks (RNNs), which may work well in learning the long-term temporal structure in music but less so for local musical textures (e.g. arpeggios and broken chords), as compared with convolutional neural networks (CNNs) \cite{musegan}.

\section{Proposed Model}
\label{sec:model}

The proposed model is composed of three stages: lead sheet generation, feature extraction, and arrangement generation, as illustrated in Figure \ref{fig:generator_arch}. We introduce them below.

\subsection{Data Representation}
\begin{figure}[t]
\centering
\includegraphics[width=.4\textwidth]{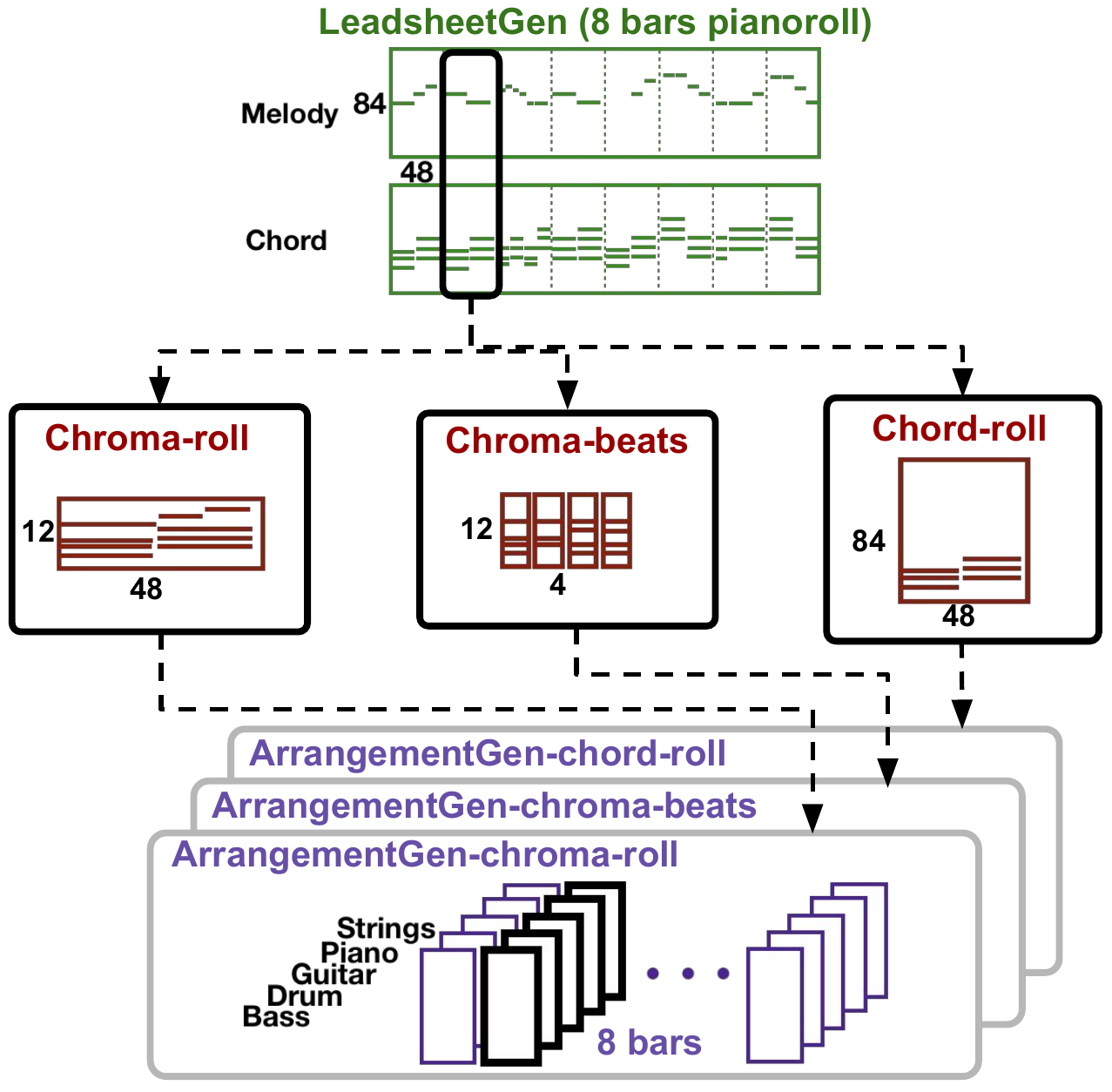}
\caption{Lead sheet arrangement system flow.}
\label{fig:system_flow}
\end{figure}

In order to model multi-track music, we adopt the piano-roll representation as our data format. A piano-roll is represented as a binary-valued score-like matrix  \cite{musegan}. Its x-axis and y-axis denote the time steps and note pitches, respectively. An \emph{N}-track piano-roll of one bar long is represented as a tensor $X \in \{0,1\}^{T\times P\times N}$, where $T$ denotes the number of time steps in a bar and $P$ stands for number of note pitches. Both lead sheets and MIDIs can be converted to piano-rolls. For example, a lead sheet can be seen as two-track (melody, chord) piano-rolls, as illustrated in the upper part of Fig. \ref{fig:system_flow}.

\subsection{Lead Sheet Generation}

The goal of our lead sheet generation model is to \textbf{generate lead sheets of eight bars long from scratch}. As the left hand side of Figure \ref{fig:generator_arch} shows, it contains two sets of generative models: the \emph{temporal} generators $G_\text{temp}$ and the \emph{bar} generators $G_\text{bar}$.
The temporal generators are in charge of the temporal dependency across bars and their output becomes part of the input of the bar generator. The bar generators generate music one bar at a time. Since the piano-rolls for each bar of a lead sheet are two image-like matrices (one for melody, one for chords), we can use CNN-based models for $G_\text{bar}$.

In order to generate realistic music, we train the generators in an ``adversarial'' way, following the principal of GAN \cite{gan}. Specifically, in the training time, we train a CNN-based discriminator $D$ (not shown in Figure \ref{fig:generator_arch}) to distinguish between real lead sheets (from existing songs) and the output of $G_\text{bar}$. While training $D$, both $G_\text{temp}$ and $G_\text{bar}$ are held fixed, and the learning target is to \emph{minimize} the classification loss of $D$. In contrast, while training $G_\text{temp}$ and $G_\text{bar}$, we fix $D$ and the learning target is to \emph{maximize} the classification loss of $D$. As a result of this iterative mini-max training process, $G_\text{temp}$ and $G_\text{bar}$ may learn to generate realistic lead sheets. 

Following the design of MuseGAN \cite{musegan}, we use four types of input random noises $z$ for our generators, to capture time dependent/independent and track dependent/independent characteristics.
However, unlike MuseGAN, we use a two-layer RNN model instead of CNN for $G_\text{temp}$. 
Empirically we find that RNNs can better capture the repetitive patterns seen in lead sheets. 
%As we have seen the feasibility to generate multi-track polyphonic by CNN structure in MuseGAN and MidiNet model, we also apply CNN to generate polyphonic music on both lead sheet and MIDI dataset. 
 %Then, bar generator will focus on generating local patterns within each bar based on the input of bar-level feature vectors. The output of bar generator is a eight-bar two-track (Melody, Chord) piano-roll.
Although such a hybrid recurrent-convolutional design is not new (e.g. used in \cite{semircnn}), to our knowledge it is the first recurrent CNN model (RCNN) for lead sheet generation.

%Instead of fully using CNN structure in all generated model as proposed in previous two papers, we apply RNN structure in temporal generation part and remain CNN structure in bar generation part. This RCNN structure could capture the often seen repetitive patterns in pop music. 

\subsection{Arrangement Generation}

The goal of our arrangement generation model, denoted as the 
\emph{conditional bar} generators $G_\text{bar}^\text{(c)}$ on the right hand side of Figure \ref{fig:generator_arch}, 
is to \textbf{generate five-track piano-rolls of one bar long conditioning on the features extracted from the lead sheets}. 
%the \emph{conditional bar} generators for arrangement generation (stage 3).
%Note that we aim at generating eight bar long music in lead sheet and one bar long conditioned music in MIDI, so CNN structure could be suitable to applied on this fix length piano-roll generation. 
This process is also illustrated in Fig. \ref{fig:system_flow}. We generate the arrangement one bar at a time, until all the eight bars have been generated (and then concatenated). We also use the principal of GAN to train $G_\text{bar}^\text{(c)}$ along with a discriminator $D^\text{(c)}$. Both $G_\text{bar}^\text{(c)}$ and $D^\text{(c)}$ are implemented as CNNs.

As the bottom-right corner of Fig. \ref{fig:generator_arch} shows, we train a CNN-based encoder $E$ along with $G_\text{bar}^\text{(c)}$ to embed the harmonic features extracted by the middle feature extractor to the same space as the output of the intermediate hidden layers of $G_\text{bar}^\text{(c)}$ and $D^\text{(c)}$. 
%The information could be properly concatenated into ArrangementGen model. The implementation details are written in the following section.
Conditioning both $G_\text{bar}^\text{(c)}$ and $D^\text{(c)}$ empirically performs better than conditioning $G_\text{bar}^\text{(c)}$ only.

\subsection{Feature Extraction}
\label{sec:model:flow}

Given a lead sheet, one may think that we can directly use the melody line or the chord sequence in the lead sheet to condition the arrangement generation. This is, however, not feasible in practice, because few MIDIs in the training data have the melody or chord tracks specified. What we have from MIDIs are usually the multi-track piano-rolls. We need to project the lead sheets and MIDIs to the same feature space to make the conditioning possible.

%On the top of the figure presents the leadsheet generation form. Both melody and chord are turn into piano-roll with its width equals to 48 time steps per bar and height equals to 84 note pitches. In this part, we generate eight-bar lead sheet music. Then, in order 

We propose to achieve this by extracting harmonic features from lead sheets and MIDIs. In this way, arrangement generation is conditioned on the harmonic part of the lead sheet. We propose the following three symbolic-domain harmonic features. See the middle of Fig. \ref{fig:system_flow} for an illustration.

\begin{itemize}
\item \textbf{Chroma piano-roll representation (chroma-roll)}: The idea is to neglect pitch octaves and compress the pitch range of a piano-roll into chroma (twelve pitch classes)  \cite{chroma}, leading to a $12 \times 48$ matrix per bar. Such a chroma representation has been widely used in audio-domain music information retrieval (MIR) tasks such as audio-domain chord recognition and cover song identification \cite{ellis07icassp}. For a lead sheet, we compute the chroma from both the melody and chord piano-rolls and then take the union. 
For a MIDI, we do the same across the $N$ piano-rolls.
%{\color{red} For MIDIs, we first sum the piano-rolls across tracks and then take the chroma.}

%but remain 48 time steps. Then, we take the union of the notes in melody track chroma and chord track chroma to form the final chroma-roll feature, which contains the information in both tracks with a more harmony-oriented way. We assume that this feature could capture harmonic part through chroma representation while still pertaining time information. 

\item \textbf{Chroma beats representation (chroma-beats)}: From chroma-roll, we further reduce the temporal resolution by taking the average per beat (i.e. per 12 time steps), leading to a $12 \times 4$ matrix per bar. 
%which is similar to the first feature but takes the beat average on it. Chorma-beats averages the notes within one beat, which equals to 12 time steps. Therefore, it has the same pitch range as chroma piano-roll representation but has only one twelfth time resolution, which contains only 4 time steps within a bar. 
The lead sheets and MIDIs may overlap more in this feature space, but the downside is the loss of detailed temporal information.
%It is a trade-off that chroma-beats feature might be easier to learn by the machine and could be easier for combination two dataset since the pattern is simple. However, it lost the detail information on time axis.

\item \textbf{Chord piano-roll representation (chord-roll)}: Instead of using chroma features, we estimate chord labels from both lead sheets and MIDIs to increase the information of harmony. 
This is done by first synthesizing the audio file of a lead sheet (or a MIDI), and then applying an audio-domain chord recognition model for chord estimation. We use the DeepChroma model implemented in the \texttt{Madmom} library \cite{madmom} for recognizing 12 major chords and 12 minor chords for each beat and use piano-roll without compressing to chroma, yielding a $84 \times 48$ matrix per bar.
We do not use the (ground truth) chord labels provided in the lead sheets, as we want the lead sheets and MIDIs to undergo the same feature extraction process.
%Although TheoryTab already provides chord labels, in order to connect two datasets with similar chord labels, we apply Madmom's chord recognition model \cite{madmom} both on synthesized TheoryTab dataset and synthesized Lakh Piano-roll dataset so that both dataset contain same types of chord labels, which includes \textit{major} and \textit{minor} chords. Before we apply Madmom's model, we also test it on a subset of MuseScore \cite{musescore}. With the testing on 706 bars, the model achieves 84.6\% accuracy on gospel sheet music, which has closer style to pop music comparing with Jazz.
% there are more than seven types of chord label, including some common ones like, \textit{major}, \textit{minor}, \textit{diminished}, \textit{seventh}, \textit{major seventh}, \textit{minor seventh} and \textit{diminished seventh} chord. Since we want to connect two dataset based on the chord labels, 
\end{itemize}

\section{Implementation}
\label{sec:impl}
This section presents the datasets used in our implementation and some technical details.
%its preprocessing are described. We also share some insights of our model settings and training process.

% \usepackage{booktabs}

\subsection{Dataset}
%There are three datasets being used in this paper and each of them has its own form, therefore, we describe the preprocessing separately.

%Many recent works has labored on generating music in lead sheet or MIDI form since there are quite sufficient amount of data in these two domains. The challenge of realizing lead sheet arrangement can be seen in the limitation of datasets. There are three useful and often used datasets listed in Table \ref{table:Datasets} 

We use the TheoryTab dataset \cite{theorytab} for the lead sheets and the Lakh piano-roll dataset \cite{musegan} for the multi-track piano-rolls. We summarize the two datasets in Table \ref{table:Datasets} and present below how we preprocess them for the sake of model training.

\emph{1) \textbf{TheoryTab Dataset (TTD)}} contains 16K segments of lead sheets stored with XML format. Since it uses scale degree to represent the chord labels, we could think of all the songs as C key songs without further key transposition. 
%Besides, there are genre tags for marking out different styles of songs. 
%For phrase-level lead sheet generation, we adopt TheoryTab dataset. 
We parse each XML file and turn the melody and chords into two piano-rolls. For each bar, we set the height to 84 (note pitch range from \texttt{C1} to \texttt{B7}) and the width (time resolution) to 48 for modeling temporal patterns such as triplets and 16th notes. As we want to generate eight-bar lead sheets, the size of the target output tensor for lead sheet generation is 8 (bars) $\times$ 48 (time steps) $\times$ 84 (notes) $\times$ 2 (tracks). We use all the songs in TTD, without filtering the songs by genre. For segments that are longer than eight bars, we take the maximum multiples of eight bars.

\emph{2) \textbf{Lakh Piano-roll Dataset (LPD)}} is derived from the Lakh MIDI dataset \cite{raffel16phd}. We use the lpd-5-cleansed subset \cite{musegan}, which contains 21,425 five-track piano-rolls that are tagged as Rock songs and are in 4/4 time signature. These five tracks are, again, strings, piano, guitar, drum and bass. Since the songs are in various keys, we transpose all the songs into C key by using the \texttt{pretty\_midi} library \cite{pretty_midi}. Since arrangement generation aims at creating arrangement of one bar at a time, the size of the target output tensor for arrangement generation is 1 (bar) $\times$ 48 (time steps) $\times$ 84 (notes) $\times$ 5 (tracks).

\begin{table}[t]
\tabcolsep=3pt
\renewcommand{\arraystretch}{1.3}
\caption{Comparison of the two datasets employed in our work. They are respectively from https://www.hooktheory.com/theorytab, and https://salu133445.github.io/lakh-pianoroll-dataset/.}
\label{table:Datasets} 
\centering
\begin{tabular}{|l|ll|}
\hline
& \textbf{TheoryTab} \cite{theorytab} & \textbf{Lakh Pianoroll} \cite{musegan} \\
\hline
\textbf{Song length}  & Segment  & Full song \\
%\hline
\textbf{Symbolic data}   & Lead sheet  & Multi-track piano-rolls \\
%\hline
\textbf{Musical key}         & C key only & Various keys  \\
%\hline
\textbf{Genre tag}   & Yes      & Yes    \\
%\hline
\textbf{Number of songs~} & 16K   & 21K  \\
\hline
\end{tabular}
\end{table}

% \begin{table}[t]
% \tabcolsep=3pt
% \renewcommand{\arraystretch}{1.3}
% \caption{Comparison of three datasets for symbolic domain music generation. The three datasets are from https://www.hooktheory.com/theorytab, http://colinraffel.com/projects/lmd/, and https://musescore.com/, respectively.}
% \label{table:Datasets} 
% \centering
% \begin{tabular}{|l|lll|}
% \hline
% & \textbf{TheoryTab} & \textbf{Lakh MIDI (LMD)} & \textbf{MuseScore}\\
% \hline
% \textbf{Length}  & Segment  & Full song & Full song\\
% %\hline
% \textbf{Symbolic data}   & Lead sheet  & MIDI & Score\\
% %\hline
% \textbf{Audio data}  & Real audio  & Real audio  & Synthesized audio\\
% %\hline
% \textbf{Key}         & C key  & Various keys  & Various keys\\
% %\hline
% \textbf{Genre tag}   & Yes      & Yes    & No \\
% %\hline
% \textbf{Number of songs} & 16K   & 176K  & 350K\\
% \hline
% \end{tabular}
% \end{table}

\subsection{Model Parameter Settings}
The $G_\text{temp}$ in the lead sheet generation model is implemented by two-layer RNN with 4 outputs and 32 hidden units. $G_\text{bar}$, $G_\text{bar}^\text{(c)}$, $D$ and $D^\text{(c)}$ are all implemented as CNNs. The total size of the input random vectors $z$ for $G_{bar}$ and $G_\text{bar}^\text{(c)}$ are both set to 128. 
%\subsection{Feature Condition}
For the encoder $E$ in arrangement generation, we adopt skip connections (as done in \cite{musegan}) and use slightly different topology for encoding the three features described in Section \ref{sec:model:flow}. 
%three types of encoder to encode the extracted feature information. 
More details of the network topology will be presented in an online appendix.
%We design the pairs of $E$, $G_\text{bar}^\text{(c)}$ and $D^\text{(c)}$ to have the same dimension in certain layers. Therefore, the harmonic information from the layers of encoder could be concatenated into corresponding layers of $G_\text{bar}^\text{(c)}$ and $D^\text{(c)}$. 
We use WGAN-gp \cite{improvewgan} for model training.
Each model is trained with a Tesla K80m GPU in less than 24 hours with batch size being 64.
\begin{table*}[t]
\renewcommand{\arraystretch}{0.99}
\centering
\caption{Result of objective evaluation for lead sheet generation, in four metrics. The values are better when they are closer to that computed from the training data (i.e., the TheoryTab dataset), shown in the first row.}
\label{table:leadsheet_eval}
\begin{tabular}{|r|cc|cc|cc|c|}
\hline
\multicolumn{1}{|c|}{{\color[HTML]{333333} }} & \multicolumn{2}{c|}{{\color[HTML]{333333} empty bars (EB)}} & \multicolumn{2}{c|}{used pitch classes (UPC)} & \multicolumn{2}{c|}{qualified notes (QN)} & tonal distance (TD) \\
\multicolumn{1}{|c|}{} & Melody & Chord & Melody & Chord & Melody & Chord & Melody-Chord \\ \hline \hline
Training data & 0.02 & 0.01 & 4.18 & 6.12 & 1.00 & 1.00 & 1.50 \\ \hline \hline
1st iteration & 0.00 & 0.00 & 16.5 & 83.7 & 0.16 & 0.42 & 0.45 \\
1,000th iteration & 0.00 & 0.00 & 6.99 & 7.59 & 0.55 & 0.71 & 1.34 \\
2,000th iteration & 0.00 & 0.00 & 6.47 & 7.85 & 0.71 & 0.80 & 1.42 \\
3,000th iteration & 0.00 & 0.00 & 4.96 & 6.75 & 0.69 & 0.84 & 1.46 \\
4,000th iteration & 0.00 & 0.00 & 4.48 & 6.66 & 0.82 & 0.87 & 1.49 \\
5,000th iteration & 0.00 & 0.00 & 4.10 & 6.69 & 0.83 & 0.84 & 1.53 \\ \hline
\end{tabular}
\end{table*}

\section{Experiment}
%In this section, we evaluate the effectiveness of lead sheet generation and arrangement generation by objective metrics and a subjective listening test, respectively. 

\subsection{Lead Sheet Generation Evaluation}

We adopt the objective metrics proposed in \cite{musegan} to  evaluate lead sheet generation, using the code they shared. \emph{Empty bars} (EB) reflects the ratio of empty bars; \emph{used pitch classes} (UPC) represents the average number of pitch classes used per bar, \emph{qualified note} (QN) denotes the ratio of notes that are longer than or equal to the 16th note (i.e. low QN suggests overly fragmented music); \emph{tonal distance} (TD) \cite{tonalDist} represents the harmonicity between two given tracks. Table \ref{table:leadsheet_eval} shows that the UPC, QN and TD of the generated lead sheets get closer to those of the training data (i.e., the TheoryTab dataset) as the learning unfolds, suggesting that the model is learning the data distribution of real lead sheets. The values gradually saturate after around 5,000 training iterations (one batch per iteration). We find no empty bars in our generation result.

\subsection{Lead Sheet Arrangement User Study}

% \begin{table}[t]
% \tabcolsep=3pt
% \renewcommand{\arraystretch}{1.3}
% \caption{Vote result on the best model surveying according to three aspects, containing harmonicity, rhythmicity and overall.}
% \label{table:Vote} 
% \centering
% \begin{tabular}{|c|ccc|}
% \hline
% & \textbf{CPF} & \textbf{CSF} & \textbf{CVF}\\
% \hline
% \textbf{Harmonicity}  & 68\%  & 16\% & 16\%\\
% %\hline
% \textbf{Rhythmicity}   & 49\%  & 35\% & 16\%\\
% %\hline
% \textbf{Overall}  & 69\%  & 13\% & 18\%\\
% \hline
% \end{tabular}
% \end{table}

% \begin{table}[t]
% \tabcolsep=8pt
% \renewcommand{\arraystretch}{1.3}
% \caption{Average score on three models from three aspects, harmonicity, rhythmicity and overall rate.}
% \label{table:Score} 
% \centering
% \begin{tabular}{|c|ccc|}
% \hline
% & \textbf{Harmonicity} & \textbf{Rhythmicity} & \textbf{Overall}\\
% \hline
% \textbf{CPF}  & 3.44  & 3.43 & 3.57\\
% %\hline
% \textbf{CSF}   & 2.18  & 3.06 & 2.60\\
% %\hline
% \textbf{CVF}  & 2.36  & 2.90 & 2.62\\
% \hline
% \end{tabular}
% \end{table}

We conduct an online user study to evaluate the result of arrangement generation. In the study, we ask respondents to listen to four groups of eight-bar music phrases arranged by our models. Each group contains one lead sheet and its three kinds of arrangements based on chroma-roll, chroma-beats and chord-roll features, respectively. The lead sheet is put at the top of each group and can be viewed as the reference. We use ``bell sound" to play the melody. We play the melody and chord along with the five tracks generated by the arrangement model to show the compatibility of these two models. After listening to the music, respondents are asked to compare the three arranged versions. They are asked to vote for the best model among the three, in terms of \emph{harmonicity}, \emph{rhythmicity} and \emph{overall feeling}, respectively.
Moreover, they are asked to rate each sample according to the same three aspects. To focus on the result of arrangement generation, we use existing lead sheets from TheoryTab for this evaluation.

\begin{figure}[t]
\centering
(a)~ \includegraphics[width=.4\textwidth]{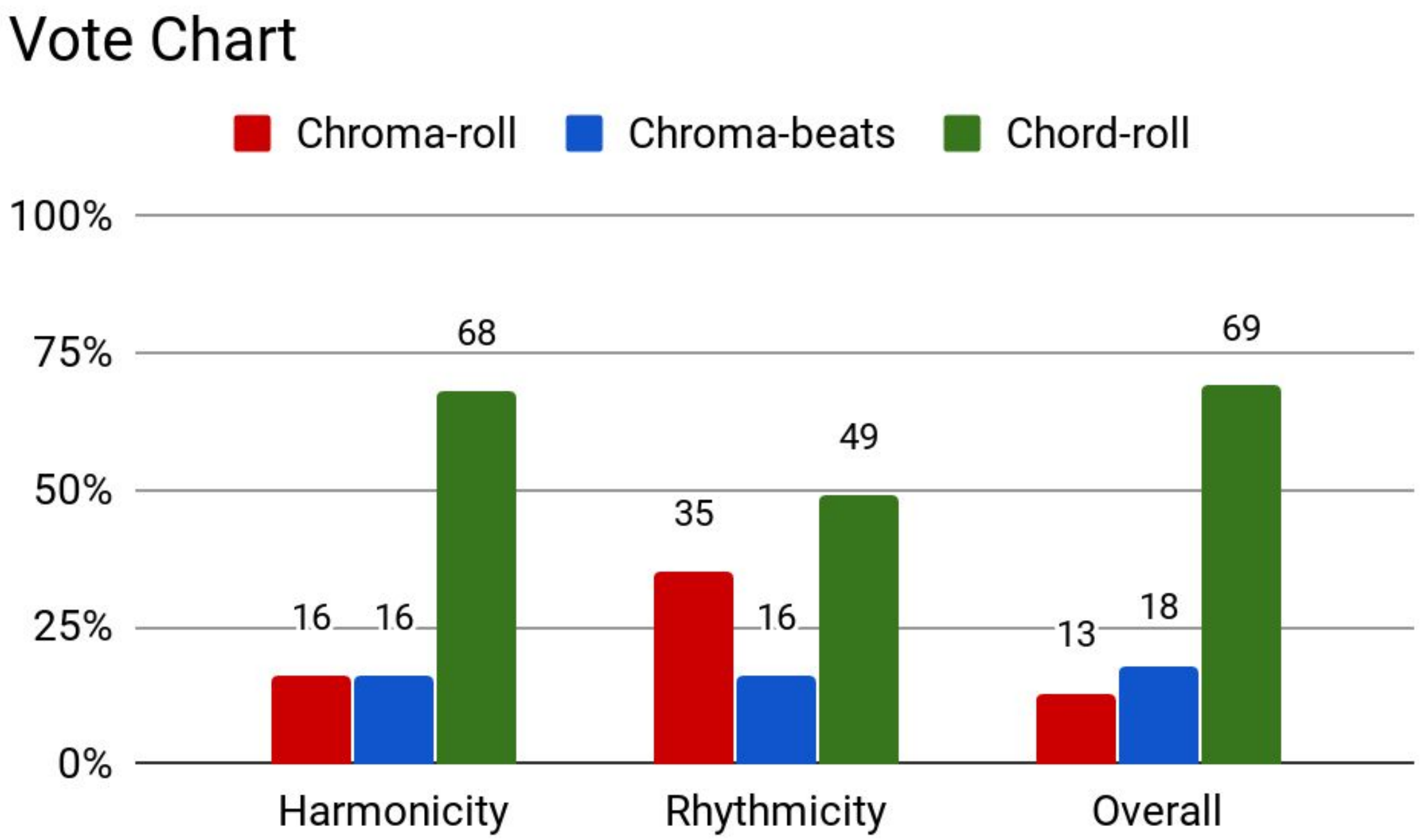}\\
(b)~ \includegraphics[width=.4\textwidth]{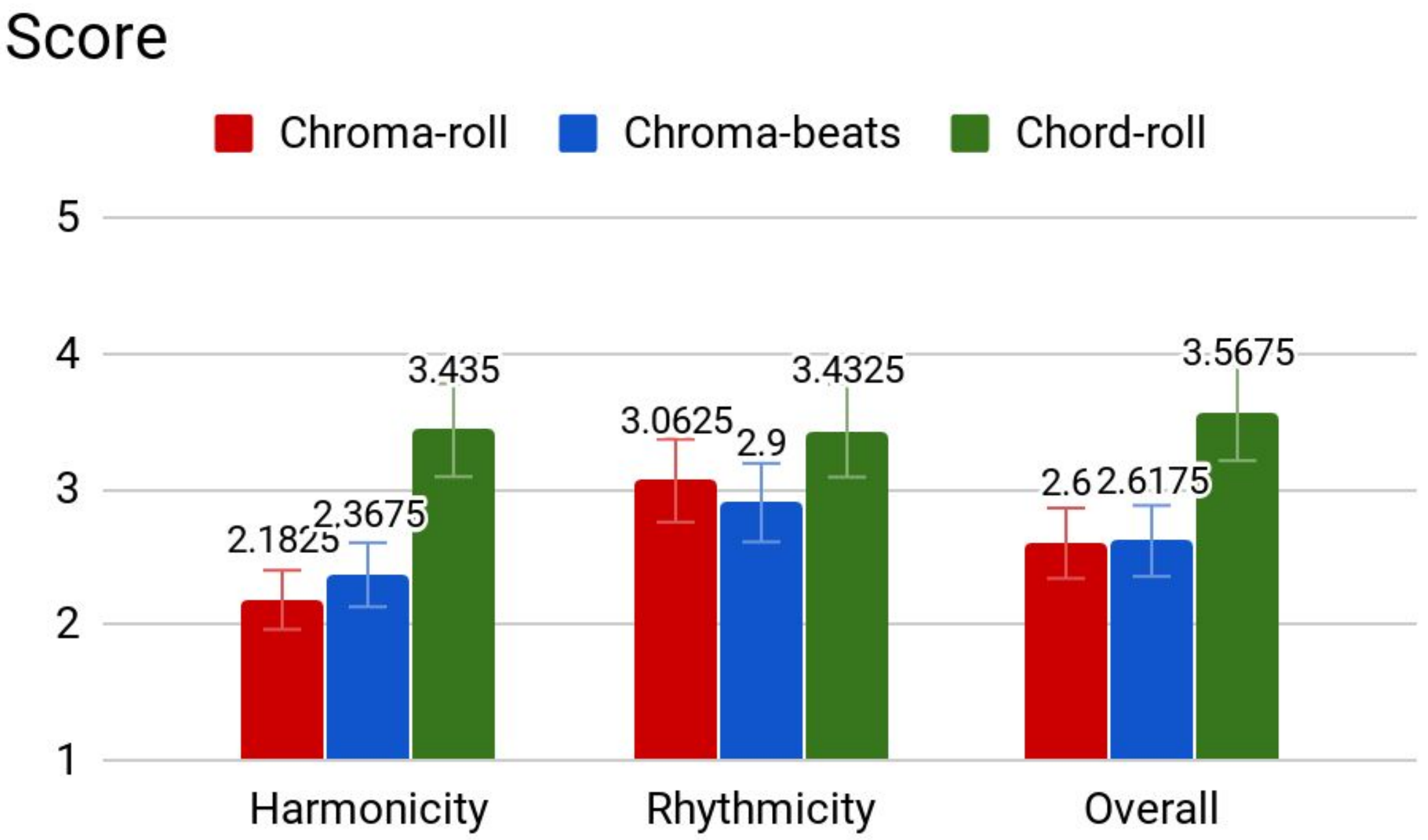}
\caption{Result of subjective evaluation for three arrangement generation models using different features, in terms of three metrics---(a) The vote result and (b) the MOS rating scores in a Likert scale from 1 to 5.}
%Vote result on the best model surveying according to three aspects, containing harmonicity, rhythmicity and overall.}
\label{fig:Vote}
\end{figure}

Figure \ref{fig:Vote} shows the average result of 25 respondents, 88\% of which play some instruments and 16\% are studying in music-related departments or working in music-related industries. The following observations can be made:
\begin{itemize}
\item In terms of harmonicity, chord-roll outperforms chroma-roll and chroma-beats by a great margin, suggesting that chords carry more harmonic information than chroma.
\item In rhythmicity, we see from the votes that chroma-beats are clearly inferior to chord-roll and chroma-roll. We attribute this to the loss of temporal resolution in the chroma-beats representation.
\item In overall feeling, chord-roll performs significantly better than the other two (which can be seen from the standard deviation shown in Fig. \ref{fig:Vote}(b)) and attains a mean opinion score (MOS) of 3.5675 in a five-point Likert scale. The result seems to suggest that harmonicity has stronger impact on the overall feeling, compared to rhythmicity. 
\end{itemize}

%In Figure \ref{fig:Vote}(a), chord-roll model achieves the largest number of vote in all three aspects. Moreover, in harmonicity and overall test, it has an obvious superiority comparing to chroma-roll model and chroma-beat model. In rhythmicity test, chroma-roll model also gains 35\% vote. However, chroma-beats model seems to get small number of vote in all three aspects.

%Figure \ref{fig:Vote}(b) shows the score for all three models with respect to harmonicity, rhythmicity and overall aspects. The score ranges from one to five. We average the score on four groups of samples and find out that chord-roll model out performs other two models in all three aspects. Moreover, these three features get similar scores on rhythmicity. We could find an interesting phenomenon that harmonicity has stronger impact on overall feeling comparing to rhythmicity. 

%In summary, chroma based feature, including chroma-roll and chroma-beats seems to have certain degree of lackness in harmonicity. The reason might be the complexed melody information is also added in the condition, which makes the model more difficult to handle the harmonic part. We can see another point from the vote chart that chroma-roll has same size as chord-roll in time resolution dimension and has 35\% vote in rhythmicity. However, chroma-beats only has one twelfth the size comparing to other two. Therefore, time resolution also influences the rhythmicity of the model.

In summary, as there are no existing work on lead sheet arrangement, we have to compare three variants of our own model. The result shows that using chord-roll to connect the lead sheets and MIDIs performs the best. The MOS in overall feeling suggests that the model is promising, but there is still room for improvement. %(as a score around 3.5 means something between okay and good).

\begin{figure}[t]
\centering
(a)~\includegraphics[width=.45\textwidth]{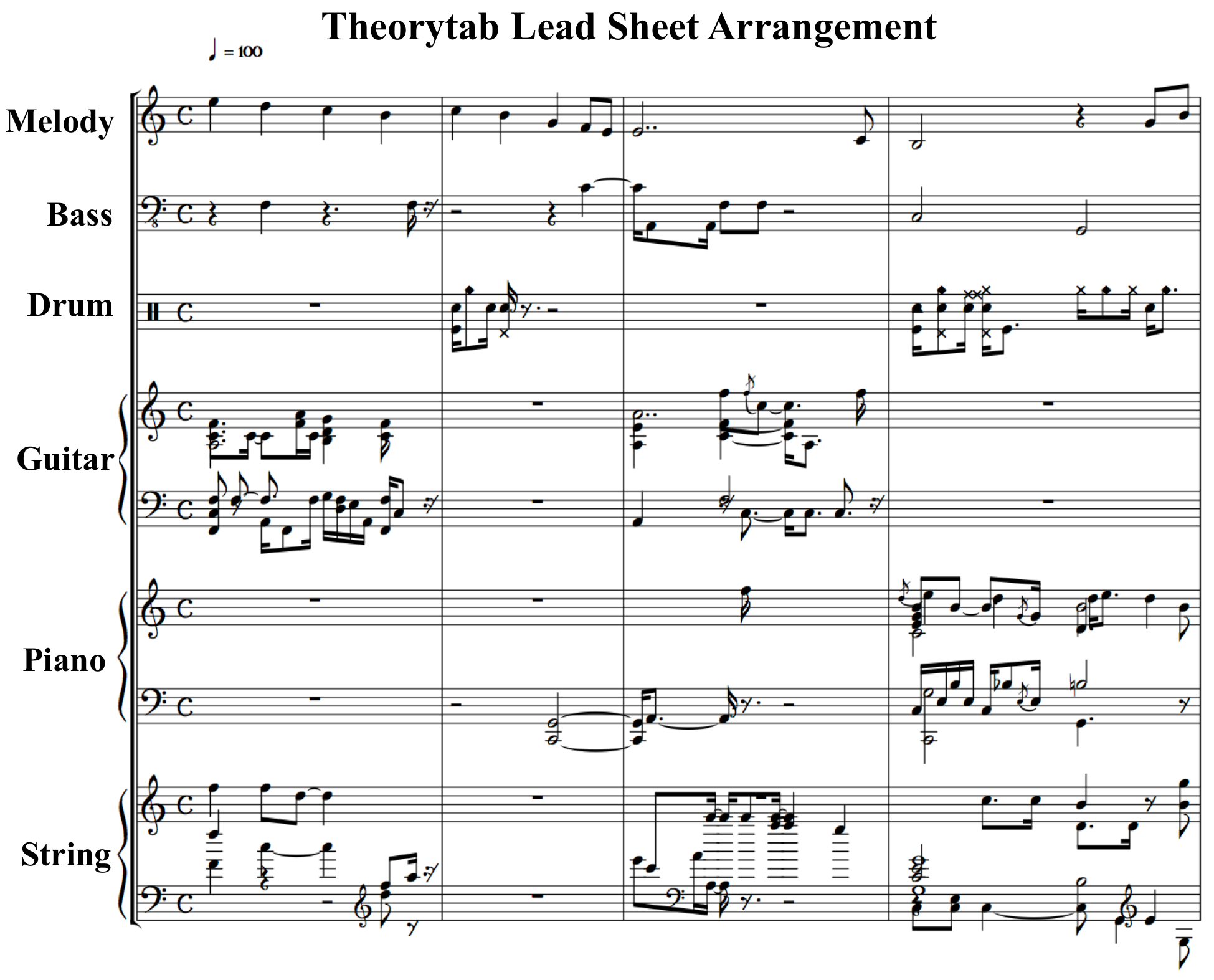} \\
(b)~\includegraphics[width=.45\textwidth]{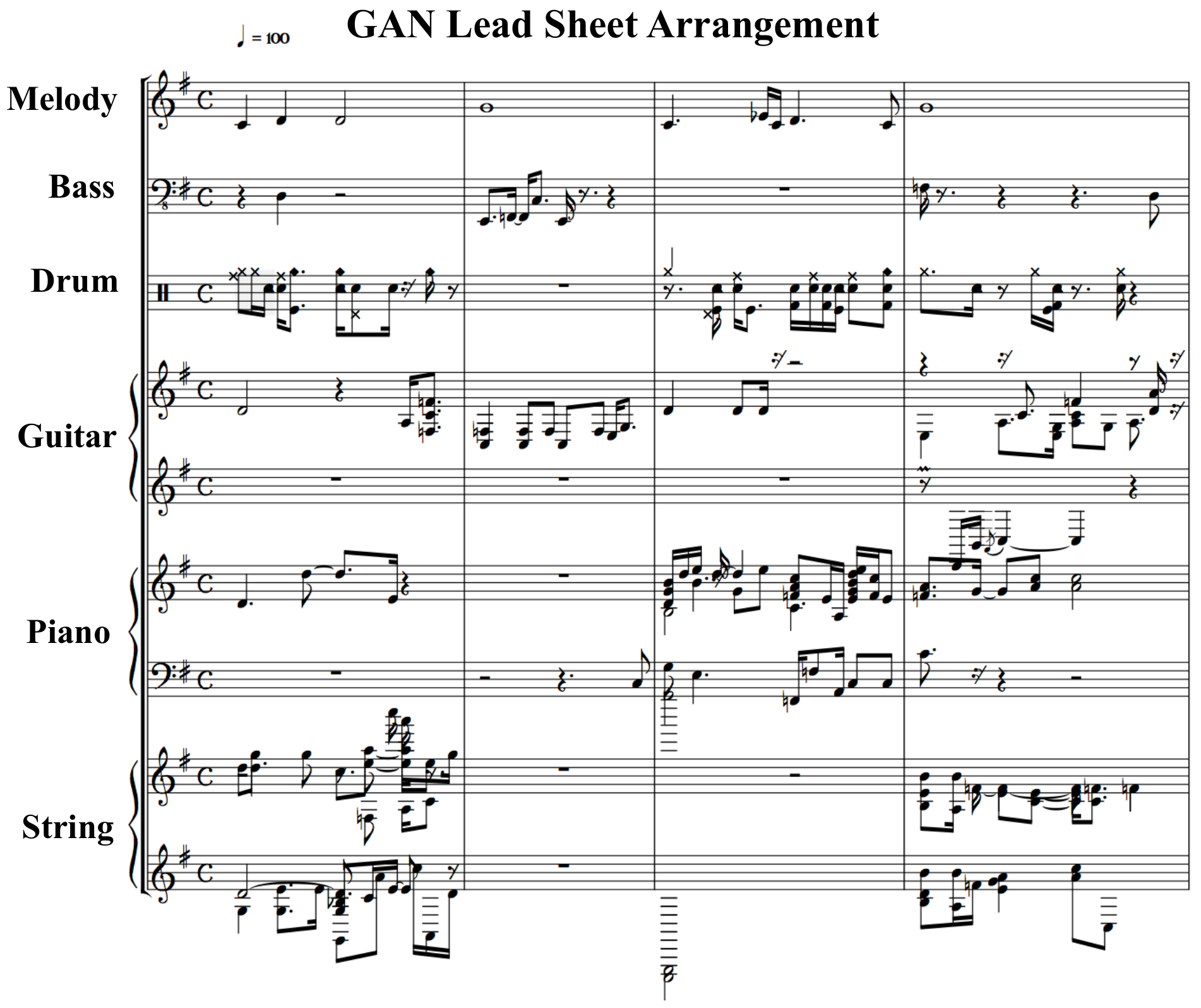}
\caption{Result of lead sheet arrangement (based on chord-roll) for (a) an existing lead sheet from TTD and (b) a lead sheet generated by our model.}
\label{fig:TT_arr_score}
\end{figure}

As illustrations, we show in Fig. \ref{fig:TT_arr_score} the scoresheet of the arranged result (based on chord-roll) of a   four-bar lead sheet from TTD and another one generated by our model. Audio examples can be found in the link described in the abstract.

% \begin{figure}[t]
% \centering
% \includegraphics[width=.47\textwidth]{./fig/GAN_arr_score.pdf}
% \caption{GAN model generated lead sheet arrangement}
% \label{fig:GAN_arr_score}
% \end{figure}

\section{Conclusion}
In this paper, we have presented a conditional GAN model for generating eight-bar phrases of lead sheets and their arrangement. 
%The music generated by this model exhibits repetitive patterns, which are often seen in Pop songs. 
To our knowledge, this represents the first model for lead sheet arrangement.
We experimented with three new harmonic features to condition the arrangement generation and found through a listening test that the chord piano-roll representation performs the best. The best model attains 3.4350, 3.4325, and 3.5675 MOS in harmonicity, rhythmicity, and overall feeling, respectively, in a Likert scale from 1 to 5.

%Moreover, we presented a conditional multi-track arrangement model, which could generate harmonic and rhythmic arrangement according to the conditional features. Finally, by comparing three proposed conditional features, chord-sequence feature shows best harmonicity while chroma-vector contains more rhythmic information. 

% \section*{Acknowledgment}

% The preferred spelling of the word ``acknowledgment'' in America is without 
% an ``e'' after the ``g''. Avoid the stilted expression ``one of us (R. B. 
% G.) thanks $\ldots$''. Instead, try ``R. B. G. thanks$\ldots$''. Put sponsor 
% acknowledgments in the unnumbered footnote on the first page.

\bibliographystyle{IEEEtran}
\bibliography{ref}

\section*{Appendix}
\begin{table}[h]
\tabcolsep=4.5pt
\renewcommand{\arraystretch}{0.7}
\caption{Conditioned generators} 
\label{table:Generator networks} 
\centering
\begin{tabular}{|lccccc|c|}
\hline
%  & Channel & Filter & Stride &  &  & output dim. \\ \hline \hline
\multicolumn{7}{|l|}{Input: $z \in \Re^{128}$} \\ \hline \hline
\multicolumn{6}{|l|}{Reshaped to (1,1) $\times$ 128 channels} & (1, 1, 128) \\
transconv\quad \space & 1024 & 1$\times$1 & (1,1) & BN & ReLU & (1, 1, 1024) \\
\multicolumn{6}{|l|}{Reshaped to (2,1) $\times$ 512 channels} & (2, 1, 512) \\
\multicolumn{6}{|l|}{chord-roll{[}5{]}} & (2, 1, 528)\\ 
transconv & 512 & 2$\times$1 & (2,1) & BN & ReLU & (4, 1, 512) \\
\multicolumn{6}{|l|}{chord-roll{[}4{]}}  & (4, 1, 528) \\
transconv & 256 & 2$\times$1 & (2,1) & BN & ReLU & (8, 1, 256) \\
\multicolumn{6}{|l|}{chord-roll{[}3{]}} & (8, 1, 272) \\
transconv & 256 & 2$\times$1 & (2,1) & BN & ReLU & (16, 1, 256) \\
\multicolumn{6}{|l|}{chord-roll{[}2{]}} & (16, 1, 272) \\
transconv & 128 & 3$\times$1 & (3,1) & BN & ReLU & (48, 1, 128) \\
\multicolumn{6}{|l|}{chord-roll{[}1{]}}  & \multicolumn{1}{l|}{(48, 1, 144)} \\ %\hline
transconv & 64 & 1$\times$7 & (1,1) & BN & ReLU & (48, 7, 64) \\
\multicolumn{6}{|l|}{chord-roll{[}0{]}}  & (48, 7, 80) \\
transconv & 1 & 1$\times$12 & (1,12) & BN & tanh & (48, 84, 1) \\ \hline \hline
\multicolumn{7}{|l|}{Output: $G_{chord-roll}(z) \in \Re^{48\times84}$} \\ \hline
\end{tabular}

\smallskip
(a) Chord-roll conditioned generator
\bigskip

\begin{tabular}{|lccccc|c|}
\hline
%  & Channel & Filter & Stride &  &  & output dim. \\ \hline \hline
\multicolumn{7}{|l|}{Input: $z \in \Re^{128}$} \\ \hline \hline
\multicolumn{6}{|l|}{Reshaped to (1,1) $\times$ 128 channels} & (1, 1, 128) \\
transconv\quad \space & 1024 & 1$\times$1 & (1,1) & BN & ReLU & (1, 1, 1024) \\
transconv & 512 & 1$\times$12 & (1,12) & BN & ReLU & (1, 12, 512) \\
\multicolumn{6}{|l|}{chroma-roll{[}5{]}} & (1, 12, 528) \\
transconv & 256 & 2$\times$1 & (2,1) & BN & ReLU & (2, 12, 256) \\
\multicolumn{6}{|l|}{chroma-roll{[}4{]}} & (2, 12, 272) \\
transconv & 256 & 2$\times$1 & (2,1) & BN & ReLU & (4, 12, 256) \\
\multicolumn{6}{|l|}{chroma-roll{[}3{]}} & (4, 12, 272) \\
transconv & 128 & 2$\times$1 & (2,1) & BN & ReLU & (8, 12, 128) \\
\multicolumn{6}{|l|}{chroma-roll{[}2{]}} & (8, 12, 144) \\
transconv & 128 & 2$\times$1 & (2,1) & BN & ReLU & (16, 12, 128) \\
\multicolumn{6}{|l|}{chroma-roll{[}1{]}} & \multicolumn{1}{l|}{(16, 12, 144)} \\
transconv & 64 & 3$\times$1 & (3,1) & BN & ReLU & (48, 12, 64) \\
\multicolumn{6}{|l|}{chroma-roll{[}0{]}} & (48, 12, 80) \\
transconv & 1 & 1$\times$7 & (1,7) & BN & tanh & (48, 84, 1) \\ \hline \hline
\multicolumn{7}{|l|}{Output: $G_{chroma-roll}(z) \in \Re^{48\times84}$} \\ \hline
\end{tabular}

\smallskip
(b) Chroma-roll conditioned generator
\bigskip

\begin{tabular}{|lccccc|c|}
\hline
%  & Channel & Filter & Stride &  &  & output dim. \\ \hline \hline
\multicolumn{7}{|l|}{Input: $z \in \Re^{128}$} \\ \hline \hline
\multicolumn{6}{|l|}{Reshaped to (1,1) $\times$ 128 channels} & (1, 1, 128) \\
transconv\quad \space & 1024 & 1$\times$1 & (1,1) & BN & ReLU & (1, 1, 1024) \\ 
transconv & 512 & 1$\times$12 & (1,12) & BN & ReLU & (1, 12, 512) \\
transconv & 256 & 2$\times$1 & (2,1) & BN & ReLU & (2, 12, 256) \\
transconv & 256 & 2$\times$1 & (2,1) & BN & ReLU & (4, 12, 256) \\
\multicolumn{6}{|l|}{chroma-beats{[}0{]}} & (4, 12, 272) \\
transconv & 128 & 2$\times$1 & (2,1) & BN & ReLU & (8, 12, 128) \\
transconv & 128 & 2$\times$1 & (2,1) & BN & ReLU & (16, 12, 128) \\
transconv & 64 & 3$\times$1 & (3,1) & BN & ReLU & (48, 12, 64) \\
transconv & 1 & 1$\times$7 & (1,7) & BN & tanh & (48, 84, 1) \\ \hline \hline
\multicolumn{7}{|l|}{Output: $G_{chroma-beats}(z) \in \Re^{48\times84}$} \\ \hline
\end{tabular}

\smallskip
(c) Chroma-beats conditioned generator
\bigskip
\end{table}

Table \ref{table:Generator networks} shows three conditioned generator networks on (a) chord-roll, (b) chroma-roll and (c) chroma-beats features, respectively. Table \ref{table:Discriminator networks} presents the discriminators (a)-(c) and encoders (d)-(f) designed for the same three features, respectively. The values showned in rows of transconv and conv (from left to right) represent: number of channels, filter size, strides, batch normalization (BN) and activation function. For fully-connected layers, the values represent (from left to right): number of hidden nodes and activation functions. LReLU stands for leaky ReLU. The column after activation function denotes the dimension of each hidden layer. The names (chord-roll, chroma-roll, chroma-beats) in generator and discriminator networks shows the skip connection on the information with respect to those feature names in the corresponding encoders.

\begin{table}[h]
\tabcolsep=4.5pt
\renewcommand{\arraystretch}{0.7}
\caption{Conditioned discriminators and encoders.}
\label{table:Discriminator networks} 
\centering
\begin{tabular}{|lcccc|c|}
\hline
%  & Channel & Filter & Stride &  & output dim. \\ \hline \hline
\multicolumn{6}{|l|}{Input: $\tilde{x} \in \Re^{1\times48\times84\times5}$} \\ \hline \hline
\multicolumn{5}{|l|}{Reshaped to (48,84) $\times$ 5 channels} & (48, 84, 5) \\
\multicolumn{5}{|l|}{chord-roll{[}6{]}} & (48, 84, 6) \\
conv\qquad \qquad \space & 128 & 1$\times$12 & (1,12)  & LReLU & (48, 7, 128 \\
conv & 128 & 1$\times$7 & (1,7)  & LReLU & (48, 1, 128) \\
conv & 128 & 2$\times$1 & (2,1)  & LReLU & (24, 1, 128) \\
conv & 128 & 2$\times$1 & (2,1)  & LReLU & (12, 1, 128) \\
conv & 256 & 4$\times$1 & (2,1)  & LReLU & (5, 1, 256) \\
conv & 512 & 3$\times$1 & (2,1)  & LReLU & (2, 1, 512) \\ \hdashline[1pt/1pt]
\multicolumn{4}{|l}{fully-connected        1024} & \multicolumn{1}{l|}{LReLU} & 1024 \\
\multicolumn{4}{|l}{fully-connected   \quad 1} & \multicolumn{1}{l|}{} & 1 \\ \hline \hline
\multicolumn{6}{|l|}{Output: $D_{chord-roll}(\tilde{x}) \in \Re$} \\ \hline
\end{tabular}

\smallskip
(a) Chord-roll conditioned discriminator
\bigskip

\begin{tabular}{|lccccc|}
\hline
%  & Channel & Filter & Stride & \multicolumn{1}{c|}{} & output dim. \\ \hline \hline
\multicolumn{5}{|l}{Input: $\tilde{x} \in \Re^{1\times48\times84\times5}$} & \multicolumn{1}{l|}{} \\ \hline \hline
\multicolumn{5}{|l|}{Reshaped to (48,84) $\times$ 5 channels} & (48, 84, 5) \\
conv\qquad \qquad \space & 128 & 1$\times$7 & (1,7) & \multicolumn{1}{c|}{LReLU} & (48, 12, 128) \\
\multicolumn{5}{|l|}{chroma-roll{[}0{]}} & (48, 12, 144) \\
conv & 128 & 3$\times$1 & (3,1) & \multicolumn{1}{c|}{LReLU} & (16, 12, 128 \\
\multicolumn{5}{|l|}{chroma-roll{[}1{]}} & (16, 12, 144) \\
conv & 128 & 2$\times$1 & (2,1) & \multicolumn{1}{c|}{LReLU} & (8, 12, 128) \\
\multicolumn{5}{|l|}{chroma-roll{[}2{]}} & (8, 12, 144) \\
conv & 128 & 2$\times$1 & (2,1) & \multicolumn{1}{c|}{LReLU} & (4, 12, 128) \\
\multicolumn{5}{|l|}{chroma-roll{[}3{]}} & (4, 12, 144) \\
conv & 256 & 2$\times$1 & (2,1) & \multicolumn{1}{c|}{LReLU} & (2, 12, 256) \\
\multicolumn{5}{|l|}{chroma-roll{[}4{]}} & (2, 12, 272) \\
conv & 512 & 2$\times$1 & (2,1) & \multicolumn{1}{c|}{LReLU} & (1, 12, 512) \\
\multicolumn{5}{|l|}{chroma-roll{[}5{]}} & (1, 12, 528) \\ \hdashline[1pt/1pt]
\multicolumn{4}{|l}{fully-connected        1024} & \multicolumn{1}{l|}{LReLU} & 1024 \\
\multicolumn{5}{|l|}{fully-connected   \quad 1} & 1 \\ \hline \hline
\multicolumn{6}{|l|}{Output: $D_{chroma-roll}(\tilde{x}) \in \Re$} \\ \hline
\end{tabular}

\smallskip
(b) Chroma-roll conditioned discriminator
\bigskip

\begin{tabular}{|lccccc|}
\hline
%  & Channel & Filter & Stride & \multicolumn{1}{c|}{} & output dim. \\ \hline \hline
\multicolumn{5}{|l}{Input: $\tilde{x} \in \Re^{1\times48\times84\times5}$} & \multicolumn{1}{l|}{} \\ \hline \hline
\multicolumn{5}{|l|}{Reshaped to (48,84) $\times$ 5 channels} & (48, 84, 5) \\
conv\qquad \qquad \space & 128 & 1$\times$7 & (1,7) & \multicolumn{1}{c|}{LReLU} & (48, 12, 128) \\
conv & 128 & 3$\times$1 & (3,1) & \multicolumn{1}{c|}{LReLU} & (16, 12, 128 \\
conv & 128 & 2$\times$1 & (2,1) & \multicolumn{1}{c|}{LReLU} & (8, 12, 128) \\
conv & 128 & 2$\times$1 & (2,1) & \multicolumn{1}{c|}{LReLU} & (4, 12, 128) \\
\multicolumn{5}{|l|}{chroma-beats{[}0{]}} & (4, 12, 144) \\
conv & 256 & 2$\times$1 & (2,1) & \multicolumn{1}{c|}{LReLU} & (2, 12, 256) \\
conv & 512 & 2$\times$1 & (2,1) & \multicolumn{1}{c|}{LReLU} & (1, 12, 512) \\ \hdashline[1pt/1pt]
\multicolumn{4}{|l}{fully-connected        1024} & \multicolumn{1}{l|}{LReLU} & 1024 \\
\multicolumn{5}{|l|}{fully-connected   \quad 1} & 1 \\ \hline \hline
\multicolumn{6}{|l|}{Output: $D_{chroma-beats}(\tilde{x}) \in \Re$} \\ \hline
\end{tabular}

\smallskip
(c) Chroma-beats conditioned discriminator
\bigskip

\begin{tabular}{|lccccc|cc|}
\hline
%  & Channel & Filter & Stride &  &  & \multicolumn{1}{c|}{output dim.} & name \\ \hline \hline
\multicolumn{7}{|l|}{Input: $y \in \Re^{48\times84}$} & \multicolumn{1}{l|}{feature name} \\ \hline \hline
\multicolumn{6}{|l|}{Reshaped to (48,84) $\times$ 1 channel} & \multicolumn{1}{c|}{(48, 84, 1)} & \multicolumn{1}{l|}{} \\
conv & 16 & 1$\times$12 & (1,12) & BN & LReLU & \multicolumn{1}{c|}{(48, 7, 16)} & chord-roll{[}0{]} \\
conv & 16 & 1$\times$7 & (1,7) & BN & LReLU & \multicolumn{1}{c|}{(48, 1, 16)} & chord-roll{[}1{]} \\
conv & 16 & 3$\times$1 & (3,1) & BN & LReLU & \multicolumn{1}{c|}{(16, 1, 16} & chord-roll{[}2{]} \\
conv & 16 & 2$\times$1 & (2,1) & BN & LReLU & \multicolumn{1}{c|}{(8, 1, 16)} & chord-roll{[}3{]} \\
conv & 16 & 2$\times$1 & (2,1) & BN & LReLU & \multicolumn{1}{c|}{(4, 1, 16)} & chord-roll{[}4{]} \\
conv & 16 & 2$\times$1 & (2,1) & BN & LReLU & \multicolumn{1}{c|}{(2, 1,16)} & chord-roll{[}5{]} \\ \hline \hline
\multicolumn{8}{|l|}{Output: chord-roll[:]} \\ \hline
\end{tabular}

\smallskip
(d) Chord-roll encoder
\bigskip

\begin{tabular}{|lccccc|cc|}
\hline
%  & Channel & Filter & Stride &  &  & \multicolumn{1}{c|}{output dim.} & name \\ \hline \hline
\multicolumn{7}{|l|}{Input: $y \in \Re^{48\times12}$} & \multicolumn{1}{c|}{feature name} \\ \hline \hline
\multicolumn{6}{|l|}{Reshaped to (48,12) $\times$ 1 channel} & \multicolumn{1}{c|}{(48, 12, 1)} & \multicolumn{1}{l|}{chroma-roll{[}0{]}} \\
conv & 16 & 3$\times$1 & (3,1) & BN & LReLU & \multicolumn{1}{c|}{(16, 12, 16)} & chroma-roll{[}1{]} \\
conv & 16 & 2$\times$1 & (2,1) & BN & LReLU & \multicolumn{1}{c|}{(8, 12, 16)} & chroma-roll{[}2{]} \\
conv & 16 & 2$\times$1 & (2,1) & BN & LReLU & \multicolumn{1}{c|}{(4, 12, 16)} & chroma-roll{[}3{]} \\
conv & 16 & 2$\times$1 & (2,1) & BN & LReLU & \multicolumn{1}{c|}{(2, 12,16)} & chroma-roll{[}4{]} \\
conv & 16 & 2$\times$1 & (2,1) & BN & LReLU & \multicolumn{1}{c|}{(1, 12, 16)} & chroma-roll{[}5{]} \\ \hline \hline
\multicolumn{8}{|l|}{Output: chroma-roll[:]} \\ \hline
\end{tabular}

\smallskip
(e) Chroma-roll encoder
\bigskip

\begin{tabular}{|lccccc|cl|}
\hline
%  & Channel & Filter & Stride &  &  & \multicolumn{1}{c|}{output dim.} & \multicolumn{1}{c|}{name} \\ \hline \hline
\multicolumn{7}{|l|}{Input: $y \in \Re^{4\times12}$} &	\multicolumn{1}{c|}{feature name} \\ \hline \hline
\multicolumn{6}{|l|}{Replicated to (4,12) $\times$ 16 channels} & \multicolumn{1}{c|}{(4, 12, 16)} & chroma-beats{[}0{]} \\ \hline \hline
\multicolumn{8}{|l|}{Output: chroma-beats[:]} \\ \hline
\end{tabular}

\smallskip
(f) Chroma-beats encoder
\bigskip

\end{table}
\end{document}